\documentclass[conference]{IEEEtran}
\IEEEoverridecommandlockouts
\usepackage{cite}
\usepackage{amsmath,amssymb,amsfonts}
\usepackage{bm}
\usepackage{array}
\usepackage{makecell}
\usepackage{booktabs}
\usepackage{algorithmic}
\usepackage{blindtext}
\usepackage{graphicx}
\usepackage{textcomp}
\usepackage[table]{xcolor}
\usepackage{subcaption}
\usepackage{bbding}
\usepackage[final]{microtype}
\def\BibTeX{{\rm B\kern-.05em{\sc i\kern-.025em b}\kern-.08em
    T\kern-.1667em\lower.7ex\hbox{E}\kern-.125emX}}
\begin{document}
\bstctlcite{IEEEexample:BSTcontrol}

\title{PACC: Propagation-Aware Channel Charting with Physics-Guided Metric Learning
% {\footnotesize \textsuperscript{*}Note: Sub-titles are not captured in Xplore and
% should not be used}

\author{Binpu~Shi, Ruihan~Li, and Min~Li}
\thanks{The authors are with the College of Information Science and Electronic Engineering, the Zhejiang Provincial Key Laboratory of Multi-Modal Communication Networks and Intelligent Information, and the National Key Laboratory of Millimeter-Wave and Terahertz Remote Sensing, Zhejiang University, Hangzhou 310027, China (e-mail: \{bp.shi, ruihan.li, min.li\}@zju.edu.cn).}
}

\maketitle

\begin{abstract}
Channel charting has emerged as a promising paradigm that maps high-dimensional channel state information into a low-dimensional latent space, facilitating tasks such as radio environment sensing and beam management. However, existing methods often rely on precise user locations or timestamp-based pseudo-labels, which are difficult to obtain in privacy-sensitive scenarios and rapidly varying wireless environments. To address these limitations, we propose propagation-aware channel charting with physics-guided metric learning (PACC), a location-free framework that constructs channel-domain supervision from propagation characteristics without requiring explicit geographic information. Specifically, PACC designs a propagation-aware dissimilarity metric that adapts to line-of-sight and non-line-of-sight propagation conditions, thereby preserving both local neighborhood relationships and the intrinsic geometry of the radio environment. Simulation results demonstrate that PACC consistently outperforms both classical dimensionality-reduction methods and state-of-the-art learning-based channel-charting approaches under diverse propagation conditions.
\end{abstract}
\begin{IEEEkeywords}
Channel charting, deep metric learning, propagation awareness, physics-guided learning.
\end{IEEEkeywords}

\section{Introduction}
Beyond-5G and sixth-generation (6G) wireless systems are expected to employ larger antenna arrays, wider bandwidths, and higher carrier frequencies, resulting in increasingly high-dimensional channel state information (CSI). Despite this complexity, CSI collected in a fixed environment is largely governed by a small set of physical factors, such as user-equipment (UE) locations and the surrounding propagation geometry \cite{studer2018channel}. Consequently, CSI samples tend to lie on a low-dimensional manifold that reflects the underlying radio environment. Learning this manifold can facilitate a wide range of wireless applications, including radio resource management, beam management, and channel prediction \cite{studer2018channel}.

Channel charting (CC) aims to recover this manifold by mapping high-dimensional CSI to a low-dimensional latent space while preserving spatial neighborhood relationships. Early CC studies have employed dimensionality-reduction techniques, such as principal component analysis (PCA), multidimensional scaling (MDS), t-distributed stochastic neighbor embedding (t-SNE), and uniform manifold approximation and projection (UMAP)~\cite{studer2018channel,stephan2024angle}. Although effective as unsupervised baselines, these methods often scale poorly and lack an explicit mapping for previously unseen CSI samples.

Deep metric learning (DML) addresses these limitations by learning a parametric charting function using architectures such as autoencoders \cite{huang2019improving}, siamese networks \cite{le2021efficient,stahlke2023indoor,stephan2024angle}, and triplet networks \cite{TripletBasedCC,euchner2022improving}. Nevertheless, their performance critically depends on the side information or dissimilarity used to define sample similarity. Existing approaches inferred neighborhood relationships from temporal proximity \cite{TripletBasedCC,euchner2022improving} or channel-domain metrics, including cosine similarity \cite{le2021efficient}, channel-impulse-response amplitudes \cite{stahlke2023indoor}, angle-delay profiles \cite{stephan2024angle}, and Doppler measurements \cite{euchner2024doppler}. Other studies incorporated geodesic modeling, uncertainty estimates, access-point locations, or time-difference-of-arrival measurements to improve chart quality or recover global coordinates \cite{euchner2024uncertainty,ahadi2025tdoa}. However, these approaches either rely on timestamps, UE motion, or employ a fixed channel-domain dissimilarity that ignores the distinct physical characteristics of line-of-sight (LoS) and non-line-of-sight (NLoS) propagation.

In practice, the physical interpretation of dominant-path parameters varies significantly with the propagation condition. Under LoS propagation, the dominant-path angle of arrival (AoA) and time of arrival (ToA) are directly related to the UE position and naturally define a geometric distance. In contrast, under NLoS propagation, the dominant path is governed by reflections or diffractions, making LoS-based geometry unreliable. Instead, the overall multipath energy distribution in the angle-delay domain provides a more robust measure of channel similarity.

Motivated by these observations, we propose propagation-aware channel charting with physics-guided metric learning (PACC), which learns channel charts without requiring UE locations or timestamp-based pseudo-labels. The main contributions of this paper are summarized as follows.
\begin{itemize}
	\item We propose a propagation-aware deep metric learning objective that constructs training triplets from CSI-derived AoA, ToA, and channel gain. The proposed loss combines AoA--ToA geometric constraints for LoS pairs with angle-delay map (ADM)-based similarity constraints for NLoS pairs, thereby preserving both neighborhood relationships and local geometric structure.
	\item We develop a channel-charting network that combines a residual convolutional backbone, gain-conditioned feature modulation, and multi-head self-attention to jointly exploit multiscale channel multipath information, including the ADM and channel gain. The resulting model provides efficient out-of-sample embedding and consistently outperforms both conventional and learning-based CC baselines under LoS- and NLoS-dominant propagation scenarios.
\end{itemize}

\emph{Notation:}
Scalars, vectors and matrices are denoted by italic letters, bold lowercase letters, and bold uppercase letters, respectively. The transpose and conjugate transpose are denoted by $(\cdot)^{\text{T}}$ and $(\cdot)^{\text{H}}$, respectively. In addition, $\mathcal{CN}(0, \sigma^2)$ represents a circularly symmetric complex Gaussian distribution with zero mean and variance $\sigma^2$.

\section{System Model and Problem Formulation}
\subsection{System Model}
We consider a single-cell uplink OFDM system, where a BS equipped with an $M$-antenna uniform linear array (ULA) serves multiple single-antenna user equipments (UEs). The uplink channel from UE $k$ to the BS on the $n$-th OFDM subcarrier is modeled as
\begin{align}
	\mathbf{h}_k[n] = \sqrt{M}\sum_{l=1}^L \alpha_l\mathbf{a}_{\mathrm{BS}}(\phi_l)e^{-j2\pi f_n (\tau_l -\tau_0)},
\end{align}
where $\alpha_l$, $\phi_l$ and $\tau_l$ denote the complex gain, AoA, and ToA of the $l$-th propagation path respectively; $f_n$ and $\tau_0$ represent the frequency of the $n$-th subcarrier and the reference delay, respectively, while $\mathbf{a}_{\text{BS}}(\cdot)$ is the BS steering vector given by
\begin{align}
	\mathbf{a}_{\text{BS}}(\phi)=\frac{1}{\sqrt{M}}[1,e^{j\pi \text{cos}(\phi)},...,e^{j(M-1)\pi \text{cos}(\phi)}]^{\text{T}}.
\end{align}

To estimate the channel, each UE transmits pilot symbol $s_k[n]$ on the $n$-th subcarrier with the transmit power $P$, i.e., $\mathbb{E}[| s_k[n] |^2 ]= P$. Assuming a digital receiver at the BS, the received signal from UE $k$ on subcarrier $n$ is expressed as
\begin{align}
	\mathbf{y}_k[n] = \mathbf{h}_k[n] s_k[n] + \mathbf{n},
\end{align}
where $\mathbf{n} \in \mathbb{C}^{M\times 1} \sim \mathcal{CN}(0, \sigma^2 \mathbf{I})$ denotes additive complex white Gaussian noise (AWGN) with variance $\sigma^2$. The corresponding least-square (LS) channel estimate is obtained as
\begin{align}
	\hat{\mathbf{h}}_k[n] = \mathbf{y}_k[n] s_k^{-1}[n].
\end{align}

Collecting the estimated channels overall $N_{\text{sub}}$ subcarriers yields the frequency-domain CSI matrix for UE $k$: 
\begin{align}
	\hat{\mathbf{H}}_k = [\hat{\mathbf{h}}_k[1], \ \hat{\mathbf{h}}_k[2],\cdots ,\hat{\mathbf{h}}_k[N_{\text{sub}}]]^{\text{T}}\in \mathbb{C}^{N_{\text{sub}}\times M}.
\end{align}

\subsection{Channel Charting}
In a quasi-static environment, the CSI observed at a fixed BS is primarily determined by the UE location and the surrounding propagation geometry. Therefore, nearby UEs tend to exhibit similar channel characteristics, and the corresponding CSI samples are expected to lie on a low-dimensional manifold that captures the underlying radio environment \cite{studer2018channel}. % \cite{studer2018channel,stephan2024angle,foliadis2024deep}.

Given a set of estimated CSI samples $\{\hat{\mathbf{H}}_k\}_{k\in\mathcal{S}}$, where $\mathcal{S}=\{1,\ldots,K\}$, channel charting thus aims to learn a forward charting function $\mathcal{Z}_{\gamma}(\cdot)$, parameterized by $\gamma$, that maps each high-dimensional CSI sample to a low-dimensional latent coordinate 
\begin{align}
	\mathcal{Z}_{\gamma}: \mathbb{C}^{N_{\text{sub}}\times M} \rightarrow \mathbb{R}^D,\qquad
	\mathcal{Z}_{\gamma}(\hat{\mathbf{H}}_k)=\mathbf{z}_k,
	\label{eq:cc_problem}
\end{align}
where $D\ll 2MN_{\text{sub}}$ is typically chosen as two or three. The objective is to preserve the spatial geometry of the radio environment, such that physically nearby UEs are mapped to nearby latent coordinates, while physically distant UEs remain well separated.

\section{PACC: Architecture and Key Designs}
This section presents the proposed PACC framework that consists of three key components: (i) extraction of propagation-aware features from CSI, including angle-delay information, channel gain and LoS/NLoS propagation classification; (ii) a deep metric learning objective that jointly preserves neighborhood relationships and enforces physics-based geometric consistency; and (iii) a neural network that fuses multiscale channel information and maps the extracted features into low-dimensional channel-chart coordinates.

\subsection{Feature Extraction}
To capture both dominant propagation paths and multipath structures, we extract three types of features from the estimated CSI for each UE: the channel gain $\hat{g}_k$, an angle-delay map (ADM) $\hat{\mathbf{\Omega}}_k$, and the dominant path's ToA and AoA.

Specifically, the overall channel gain is estimated by the Frobenius norm of the estimated CSI, i.e., $\hat{g}_k=\lVert\hat{\mathbf{H}}_k\rVert_{\mathrm{F}}$.
Additionally, two DFT-based dictionaries, $\mathbf{A}\in\mathbb{C}^{N_{\text{sub}}\times G_1}$ and $\mathbf{B}\in\mathbb{C}^{M\times G_2}$, are employed to transform the frequency-domain CSI into the delay and angular domains. The ADM is obtained as
\begin{align}
	\hat{\mathbf{\Omega}}_k = |\mathbf{A}^{\text{H}}\hat{\mathbf{H}}_k\mathbf{B}|
	\in \mathbb{R}^{G_1 \times G_2},
\end{align}
where $|\cdot|$ denotes the element-wise magnitude, and $G_1$ and $G_2$ are the numbers of delay and angular bins, respectively. The dominant propagation path is identified as
\begin{align}
	(\hat{i}_k^{(\tau)}, \hat{j}_k^{(\theta)}) = \arg\max_{(i,j)} \hat{\boldsymbol{\Omega}}_k[i,j],
\end{align}
where the corresponding physical ToA $\hat{\tau}_k$ and AoA  $\hat{\theta}_k$ are obtained from the delay and angular grids $(\hat{i}_k^{(\tau)}, \hat{j}_k^{(\theta)})$.

To remove less-informative delay components, the ADM is adaptively cropped into a fixed-size window
$\tilde{\mathbf{\Omega}}_k \in \mathbb{R}^{N_\tau \times G_2}$ around the dominant path. Specifically, the starting delay index is set as  $\tau_{\min,k}=\hat{i}_k^{(\tau)}-\tau_w$ 
with boundary adjustment when necessary. This operation retains the dominant path and nearby multipath components while suppressing weak and less-informative components.

Moreover, since the geometric interpretation of propagation paths differs between LoS and NLoS conditions, we identify the propagation condition of each CSI sample. Specifically, the root-mean-square (RMS) delay spread is extracted from the ADM. By aggregating the ADM power over all angular bins, the power-delay profile (PDP) is obtained as
\begin{align}
	\mathbf{p}_k[i]=\sum_{j=1}^{G_2}\left|\hat{\mathbf{\Omega}}_k[i,j]\right|^2.
\end{align}

The corresponding RMS delay spread is calculated as
\begin{align}
	\tau_{\mathrm{RMS},k}=
	\sqrt{\frac{\sum_{i=1}^{G_1}(\tau_i-\bar{\tau}_k)^2\mathbf{p}_k[i]}
	{\sum_{i=1}^{G_1}\mathbf{p}_k[i]}},
\end{align}
where the mean excess delay is $\bar{\tau}_k=\frac{\sum_{i=1}^{G_1}\tau_i \mathbf{p}_k[i]} {\sum_{i=1}^{G_1}\mathbf{p}_k[i]}$. Since LoS channels typically exhibit a more concentrated PDP than NLoS channels, the propagation label is determined as \cite{muqaibel2025under}
\begin{align}
	s_k=
	\begin{cases}
	1, & \tau_{\mathrm{RMS},k}\leq \tau_{\mathrm{DS}}, \quad \text{LoS},\\
	0, & \tau_{\mathrm{RMS},k}>\tau_{\mathrm{DS}}, \quad \text{NLoS},
	\end{cases}
	\label{eq:losandnlos}
\end{align}
where $\tau_{\mathrm{DS}}$ is a threshold selected according to the empirical RMS delay-spread distribution of the training data.

\subsection{Loss Function Design}
The proposed DML objective jointly preserves neighborhood relationships and local geometric structure. Specifically, a triplet loss first learns the relative ordering of neighboring CSI samples, while a physics-guided loss further preserves their local geometric relationships.

For each anchor CSI sample $\hat{\mathbf{H}}_i$, a positive sample $\hat{\mathbf{H}}_j$  and a negative sample $\hat{\mathbf{H}}_k$ are selected according to their physical features. Let $\hat{\mathbf{x}}_p=(\hat{\tau}_p,\hat{\theta}_p,\hat{g}_p)^{\text{T}}$
and define a threshold vector
$\boldsymbol{\epsilon}=(\tau_{\mathrm{th}},\theta_{\mathrm{th}},g_{\mathrm{th}})^{\text{T}}$.
The training triplet set is constructed as
\begin{align}
	\mathcal{T}\subset\left\{(i,j,k)\in\mathcal{S}_{\text{train}}^3\quad\mathrm{s.t.}\quad
	\begin{aligned}
		&|\hat{\mathbf{x}}_j - \hat{\mathbf{x}}_i| \leq \boldsymbol{\epsilon} \\
		&|\hat{\mathbf{x}}_k - \hat{\mathbf{x}}_i| \nleq \boldsymbol{\epsilon}
	\end{aligned}\right\},\end{align}
where the inequalities are evaluated element-wise. Accordingly, the positive sample is similar to the anchor in terms of ToA, AoA, and channel gain, whereas the negative sample violates at least one of these criteria. The corresponding triplet loss is defined as \cite{TripletBasedCC}
\begin{multline}
    \mathcal{L}_{\text{tri}}=
    \frac{1}{|\mathcal{T}|}
    \sum_{(i,j,k)\in \mathcal{T}}
    \bigl[\lVert\mathcal{Z}_{\gamma}
    (\hat{\mathbf{H}}_i)-\mathcal{Z}_{\gamma}(\hat{\mathbf{H}}_j)\rVert \\
    - \lVert\mathcal{Z}_{\gamma}(\hat{\mathbf{H}}_i)-\mathcal{Z}_{\gamma}(\hat{\mathbf{H}}_k)\rVert+\mathcal{M}\bigr]^+,
    \label{eq:loss_triplet}
\end{multline}
where $[u]^+=\max(u,0)$, and $\mathcal{M}$ is a margin parameter. 

Although the triplet loss above preserves neighborhood ordering, it does not explicitly constrain the relative distances among neighboring samples. To preserve the local geometry of the channel chart, we further define a propagation-aware dissimilarity $d_{i,j}$ for each positive pair $(i,j)$. 

Specifically, for LoS propagation, the dominant-path ToA and AoA provide meaningful geometric information. The dissimilarity is therefore defined using the law of cosines as
\begin{align}
	d_{i,j}^{\text{(LoS)}} = \sqrt{\hat{\tau}_i^2+\hat{\tau}_j^2
	-2\hat{\tau}_i\hat{\tau}_j
	\cos(|\hat{\theta}_i-\hat{\theta}_j|)}.
\end{align}

In contrast, under NLoS propagation, the dominant path no longer directly reflects the UE geometry. However, neighboring UEs tend to exhibit similar multipath structures, leading to the dissimilarity
\begin{align}
	d_{i,j}^{\text{(NLoS)}} = \eta \ \left(1 - \frac{\text{vec}(\hat{\mathbf{\Omega}}_i)^{\text{T}} \text{vec}(\hat{\mathbf{\Omega}}_j)}{\lVert \hat{\mathbf{\Omega}}_i \rVert \lVert \hat{\mathbf{\Omega}}_j \rVert}\right),
	\label{eq:nlosL}
\end{align}
where $\operatorname{vec}(\cdot)$ denotes vectorization, and the learnable parameter $\eta$ aligns the scales of the LoS and NLoS dissimilarities. 

The overall propagation-aware dissimilarity is given by 
\begin{align}
	d_{i,j} = \sigma(i,j) d_{i,j}^{\text{(LoS)}} + (1 - \sigma(i,j)) d_{i,j}^{\text{(NLoS)}},
\end{align}
where $\sigma(i,j)=1$ if the pair is classified as LoS and 0 otherwise according to Equation (\ref{eq:losandnlos}). 

Let $\mathcal{P}=\{(i,j):(i,j,k)\in\mathcal{T}\}$ denote the set of positive pairs. The physics-guided loss is defined as 
\begin{align}
    \mathcal{L}_{\text{pg}}=\frac{1}{|\mathcal{P}|}
    \sum_{(i,j)\in \mathcal{P}}
    \left(\lVert\mathcal{Z}_{\gamma}(\hat{\mathbf{H}}_i)
    -\mathcal{Z}_{\gamma}(\hat{\mathbf{H}}_j)\rVert-d_{i,j}\right)^2.
\end{align}

Finally, the overall training objective is
\begin{align}
	\mathcal{L}=\lambda_1 \mathcal{L}_{\text{tri}}+ \lambda_2 \mathcal{L}_{\text{pg}},
\end{align}
where parameters $\lambda_1\geq 0$ and $\lambda_2\geq 0$ balance neighborhood discrimination and local geometry preservation.

\subsection{Proposed Charting Network Architecture}
\begin{figure}[t]
	\centering
	\includegraphics[width=0.85\linewidth]{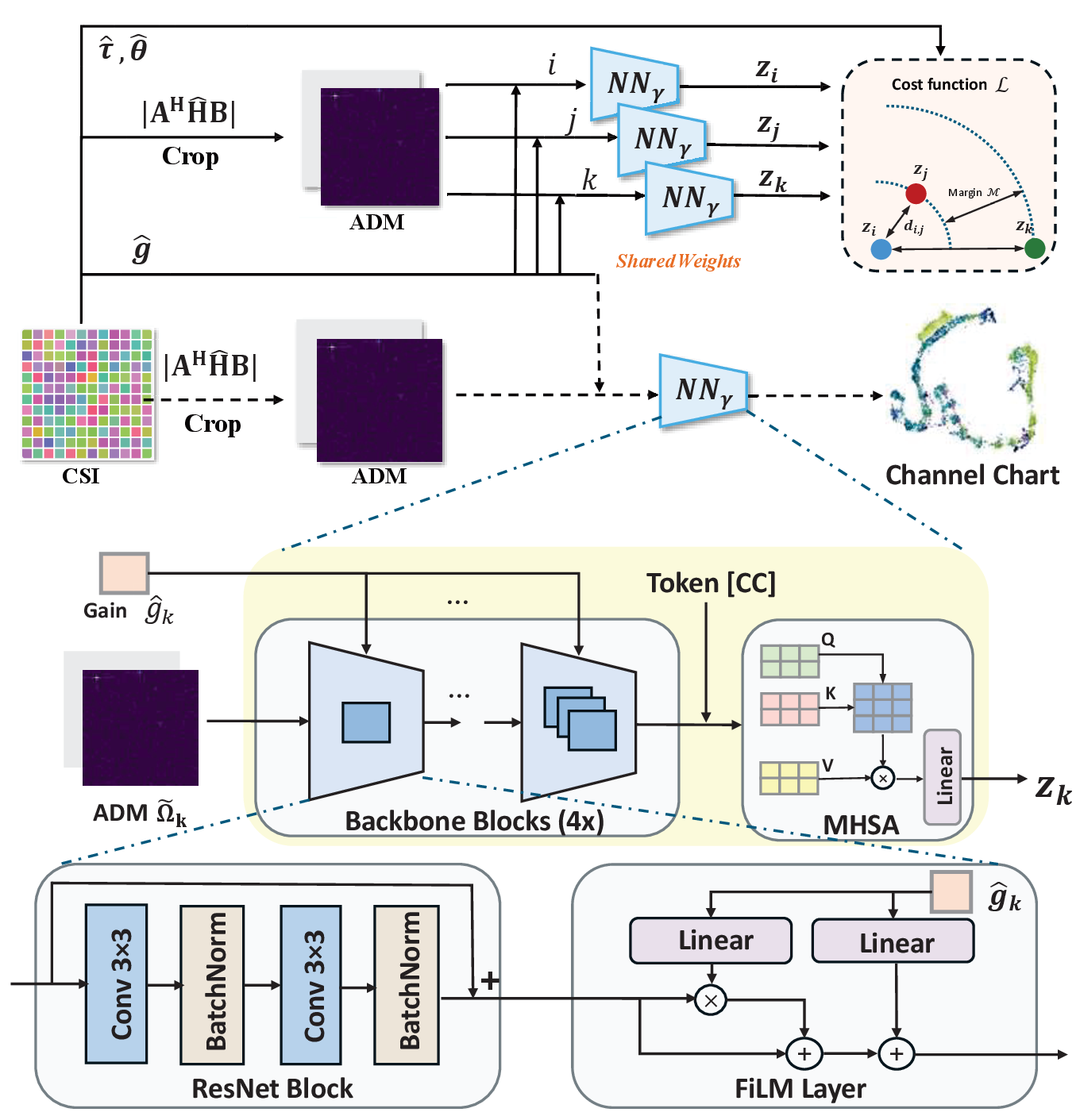}
	\caption{Architecture of the proposed PACC triplet framework.}
	\label{fig:flow2}
\end{figure} 
As illustrated in Fig.~\ref{fig:flow2}, the proposed charting network maps the normalized cropped ADM $\tilde{\mathbf{\Omega}}_k$ and channel gain $\hat{g}_k$ to the channel-chart coordinate $\mathbf{z}_k$. The ADM provides a sparse representation of the propagation environment, where dominant and reflected paths appear as localized responses in the angle-delay domain, while the channel gain captures complementary large-scale propagation information. Accordingly, the network is designed to jointly exploit local multipath patterns, channel strength, and global dependencies among propagation components.

The backbone consists of four cascaded residual blocks. Each block contains two $3\times3$ convolutional layers with a residual connection to extract local propagation patterns from adjacent angle-delay bins while facilitating stable deep feature learning. This hierarchical architecture progressively captures increasingly abstract multipath representations while preserving weak but informative propagation components \cite{he2016deep}.

To incorporate channel gain, each residual block is equipped with a gain-conditioned feature-wise linear modulation (FiLM) layer \cite{perez2018film}. Specifically, two learnable transformations map the channel gain $\hat{g}_k$ to a channel-wise scaling vector $\boldsymbol{\gamma}_k$ and bias vector $\boldsymbol{\beta}_k$. The residual feature map is then modulated as
\begin{align}
    \mathbf{F}_{\mathrm{out}}=(1+\boldsymbol{\gamma}_k)\odot
    \mathbf{F}_{\mathrm{Res}}+\boldsymbol{\beta}_k,
\end{align}
where $\odot$ denotes element-wise multiplication. By conditioning intermediate feature maps rather than only the final embedding, FiLM enables the network to adapt its feature extraction to the received channel strength throughout the backbone.

The final feature map is flattened into a sequence of angle-delay tokens, to which a learnable [CC] token is prepended. The resulting sequence is processed by a multi-head self-attention module, where each attention head computes
\begin{align}
    \operatorname{Attn}(\mathbf{Q},\mathbf{K},\mathbf{V})
    =\operatorname{softmax}\left(\frac{\mathbf{Q}\mathbf{K}^{\text{T}}}{\sqrt{d_h}}\right)\mathbf{V},
\end{align}
where $d_h$ is the dimension of each attention head. Unlike convolutional layers, which primarily capture local interactions, self-attention models long-range dependencies among propagation components in the angle-delay domain. 
A learnable [CC] token is prepended to the input token sequence to aggregate information from all tokens through the self-attention mechanism, thereby yielding a compact channel representation, which is finally projected to the $D$-dimensional coordinate $\mathbf{z}_k$. Overall, the proposed architecture combines hierarchical multipath feature extraction, gain-conditioned modulation, and global dependency modeling, making it well suited for learning compact channel-chart representations from CSI.

\section{Simulation Results}
\subsection{Dataset and Simulation Setup}
\begin{figure}[t]
	\centering
	% 第一个子图
	\begin{subfigure}{0.40\linewidth}
		\centering
		\includegraphics[width=\linewidth]{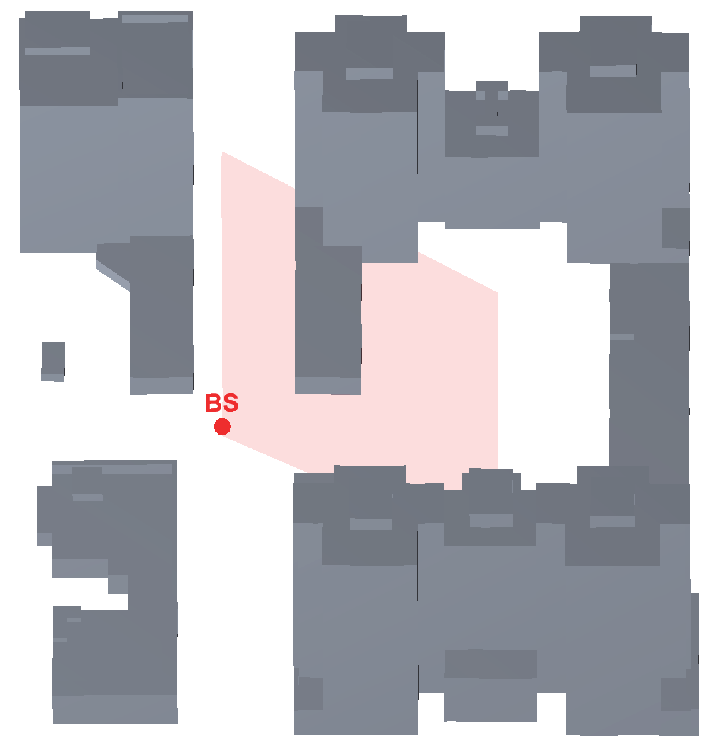}
		\caption{Scenario 1}
		\label{fig:sub-a}
	\end{subfigure}
	\quad
	% 第二个子图
	\begin{subfigure}{0.45\linewidth}
		\centering
		\includegraphics[width=\linewidth]{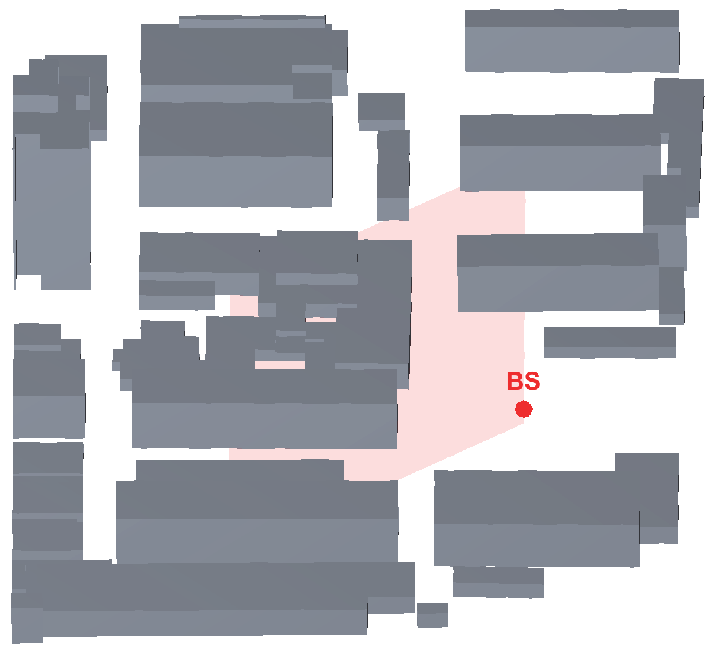}
		\caption{Scenario 2}
		\label{fig:sub-b}
	\end{subfigure}
	\caption{Wireless InSite ray-tracing environments: (a) LoS-dominant Scenario~1 and (b) NLoS-dominant Scenario~2. The red marker indicates the BS location, and the shaded area represents the UE sampling region.}
	\label{fig:Beijing}
\end{figure}

We evaluate the proposed PACC framework using channel data generated by the Wireless InSite ray-tracing simulator. As shown in Fig.~\ref{fig:Beijing}, two urban propagation scenarios are considered: Scenario~1, dominated by LoS propagation, and Scenario~2, with more severe NLoS propagation. In both scenarios, the BS is deployed at a height of $15$~m, while the UEs are randomly distributed at a height of $1$~m within a diamond-shaped region of approximately $75$~m width. The BS employs a ULA whose broadside points toward the center of the UE region at an azimuth angle of $120^{\circ}$.

For each scenario, $12{,}000$ CSI samples are generated, of which $8{,}000$ samples are used for training and $4{,}000$ for testing. The BS is equipped with $M=64$ antennas, and the OFDM system employs $N_{\text{sub}}=256$ subcarriers. The carrier frequency and bandwidth are set to $f_c=28$~GHz and $W=100$~MHz, respectively. The uplink pilot power is fixed at $P=30$~dBm, and receiver noise power is calculated as $\sigma^2=-174+10\lg W$~dBm. The delay-spread threshold \(\tau_{\mathrm{DS}}\) is determined from the training-set frequency distribution by selecting the local minimum between its two peaks~\cite{muqaibel2025under}, and we set \(\lambda_1=\lambda_2=1\). In addition, all methods produce a two-dimensional channel chart ($D=2$). 

We compare PACC with seven representative baselines, including four classical dimensionality-reduction methods (PCA, MDS, UMAP, and t-SNE) \cite{studer2018channel,stephan2024angle}, and three learning-based CC approaches: AE, Geodesic Siamese Network (GSN)~\cite{stahlke2023indoor}, and ADP~\cite{stephan2024angle}. To evaluate the contribution of the proposed physics-guided constraint, we also consider a triplet-only variant, denoted by PACC-$\mathcal{L}_{\mathrm{tri}}$. For all learning-based methods, the batch size, initial learning rate, and training epochs are set to 128, $10^{-3}$, and 120, respectively, with cosine annealing used for learning-rate scheduling.

\subsection{Numerical Results}
Following standard channel-charting evaluation criterion \cite{le2021efficient,stahlke2023indoor,stephan2024angle,TripletBasedCC,euchner2022improving}, we evaluate both neighborhood preservation and global geometric consistency. Specifically, trustworthiness (TW) and continuity (CT) measure local neighborhood preservation, while Kruskal's stress (KS) and Rajski's distance (RD) quantify the discrepancy between pairwise distances in the learned chart and the physical space \cite{venna2001neighborhood, kruskal1964multidimensional}. Higher TW/CT and lower KS/RD indicate better chart quality. 

\par
\begin{figure*}[t]
	\centering
	\begin{subfigure}[t]{0.19\textwidth}
		\centering
		\includegraphics[width=\linewidth,height=0.11\textheight,keepaspectratio]{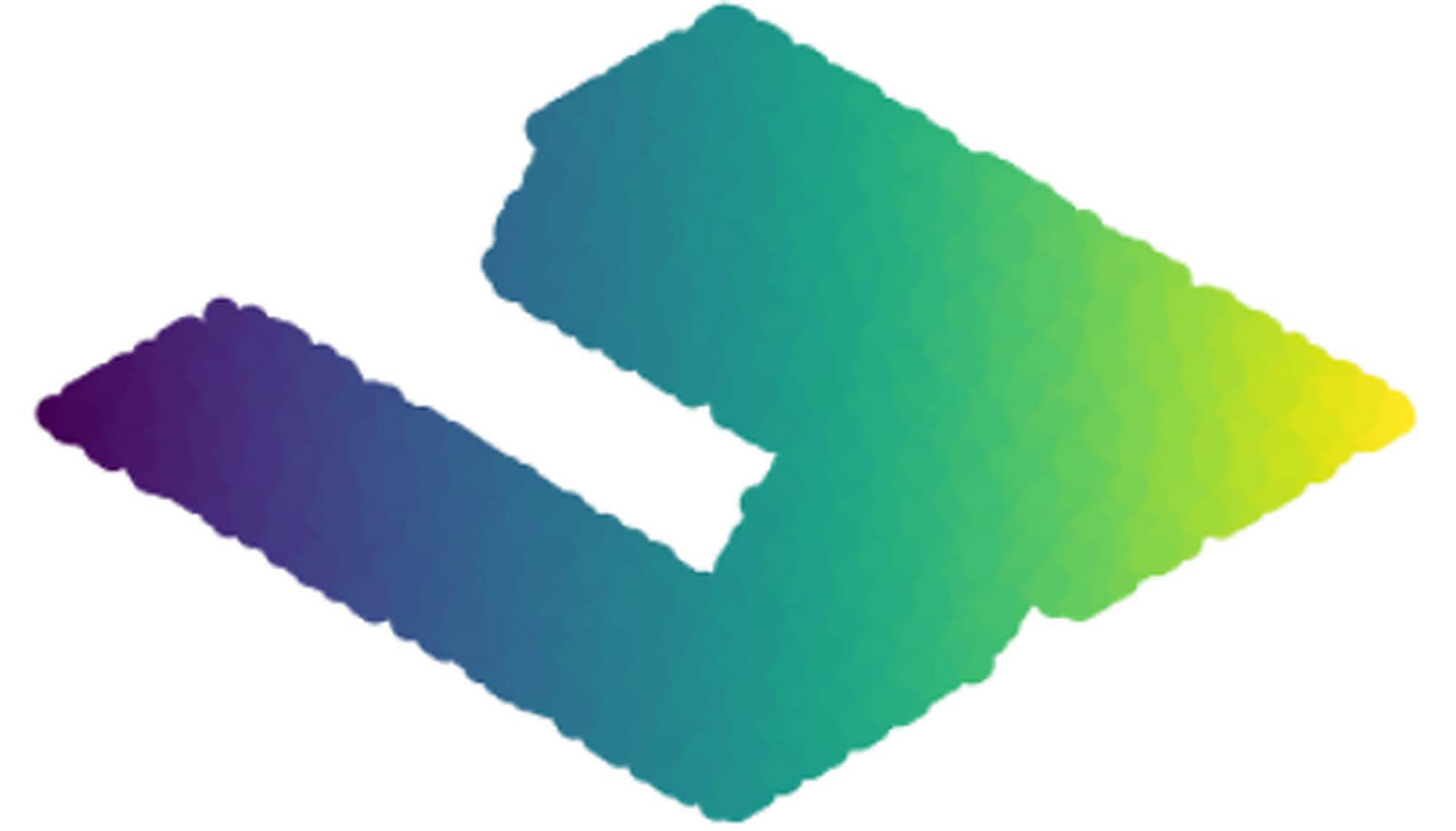}
		\caption{Real locations}
		\label{fig:scenario1_geographic}
	\end{subfigure}\hfill
	\begin{subfigure}[t]{0.19\textwidth}
		\centering
		\includegraphics[width=\linewidth,height=0.11\textheight,keepaspectratio]{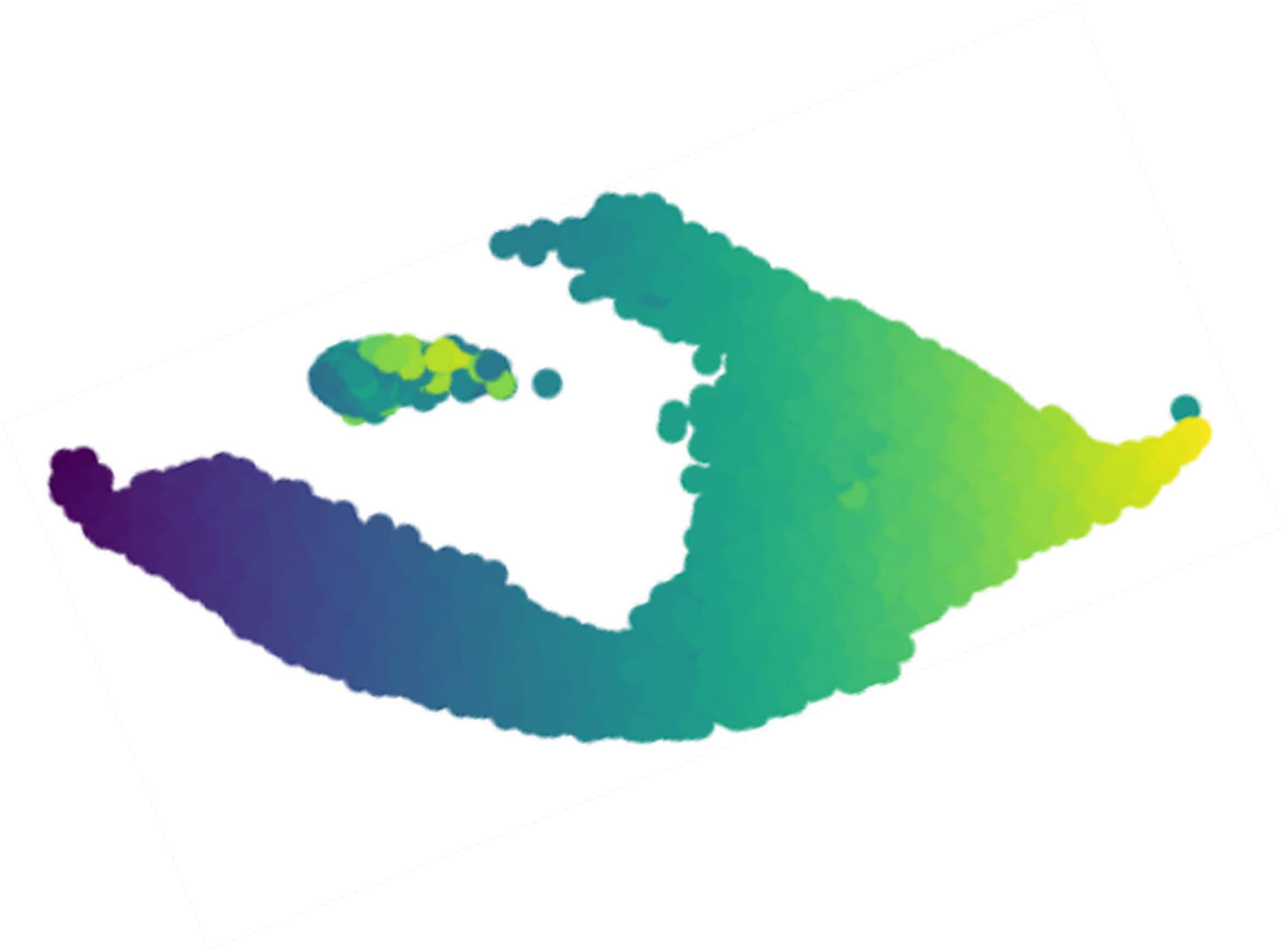}
		\caption{PACC}
		\label{fig:scenario1_pacc}
	\end{subfigure}\hfill
	\begin{subfigure}[t]{0.19\textwidth}
		\centering
		\includegraphics[width=\linewidth,height=0.11\textheight,keepaspectratio]{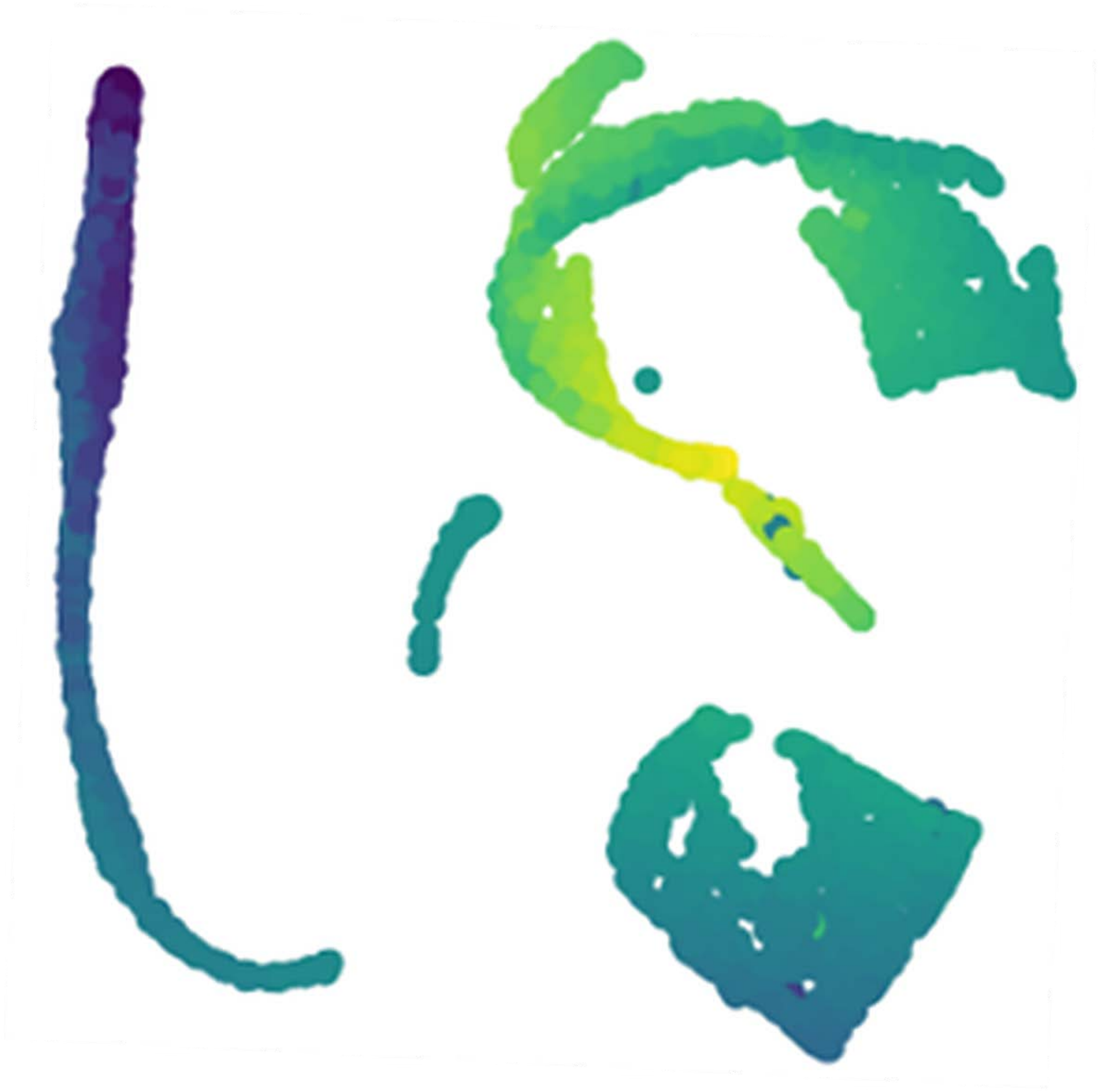}
		\caption{UMAP}
		\label{fig:scenario1_umap}
	\end{subfigure}\hfill
	\begin{subfigure}[t]{0.19\textwidth}
		\centering
		\includegraphics[width=\linewidth,height=0.11\textheight,keepaspectratio]{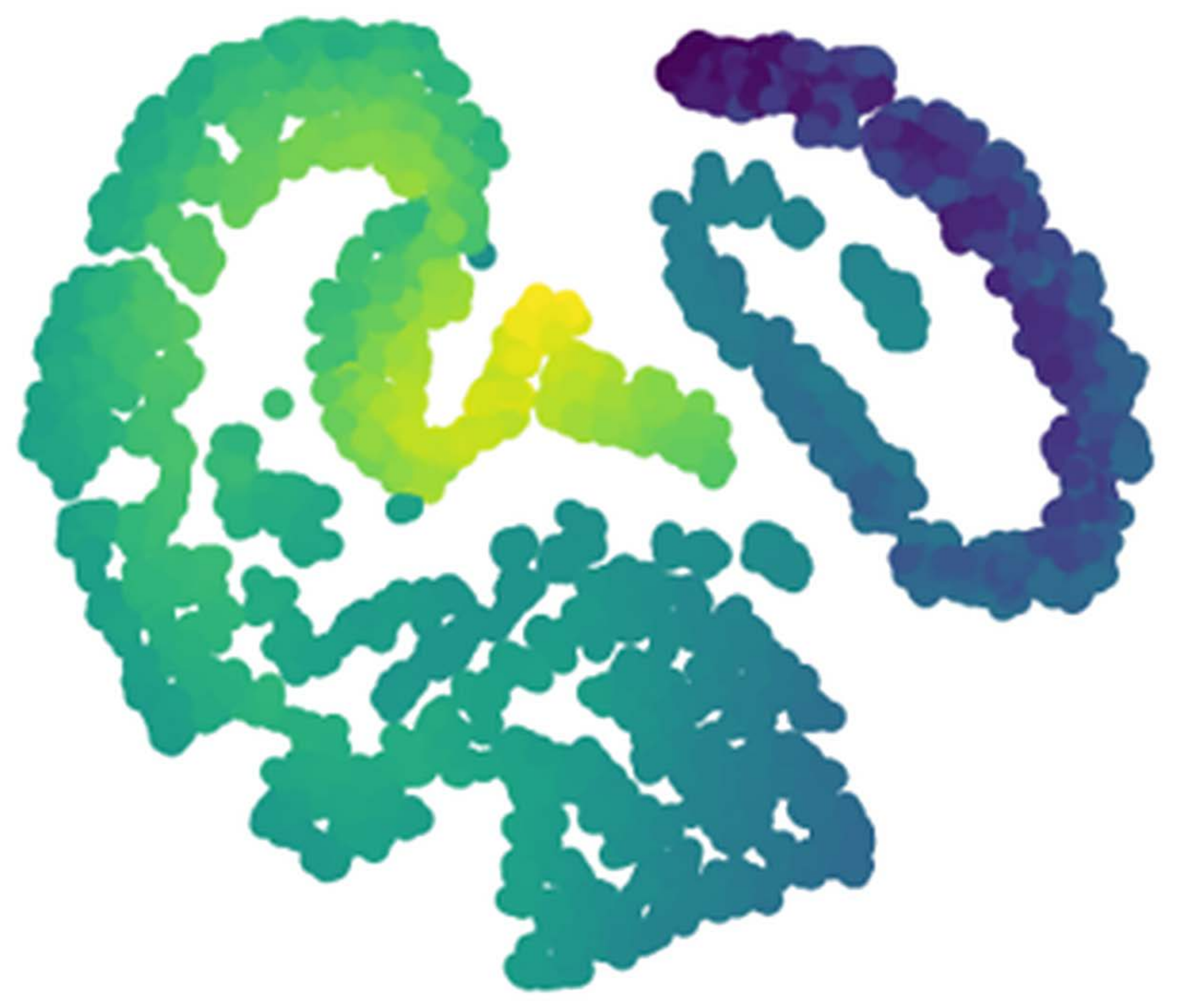}
		\caption{t-SNE}
		\label{fig:scenario1_tsne}
	\end{subfigure}\hfill
	\begin{subfigure}[t]{0.19\textwidth}
		\centering
		\includegraphics[width=\linewidth,height=0.11\textheight,keepaspectratio]{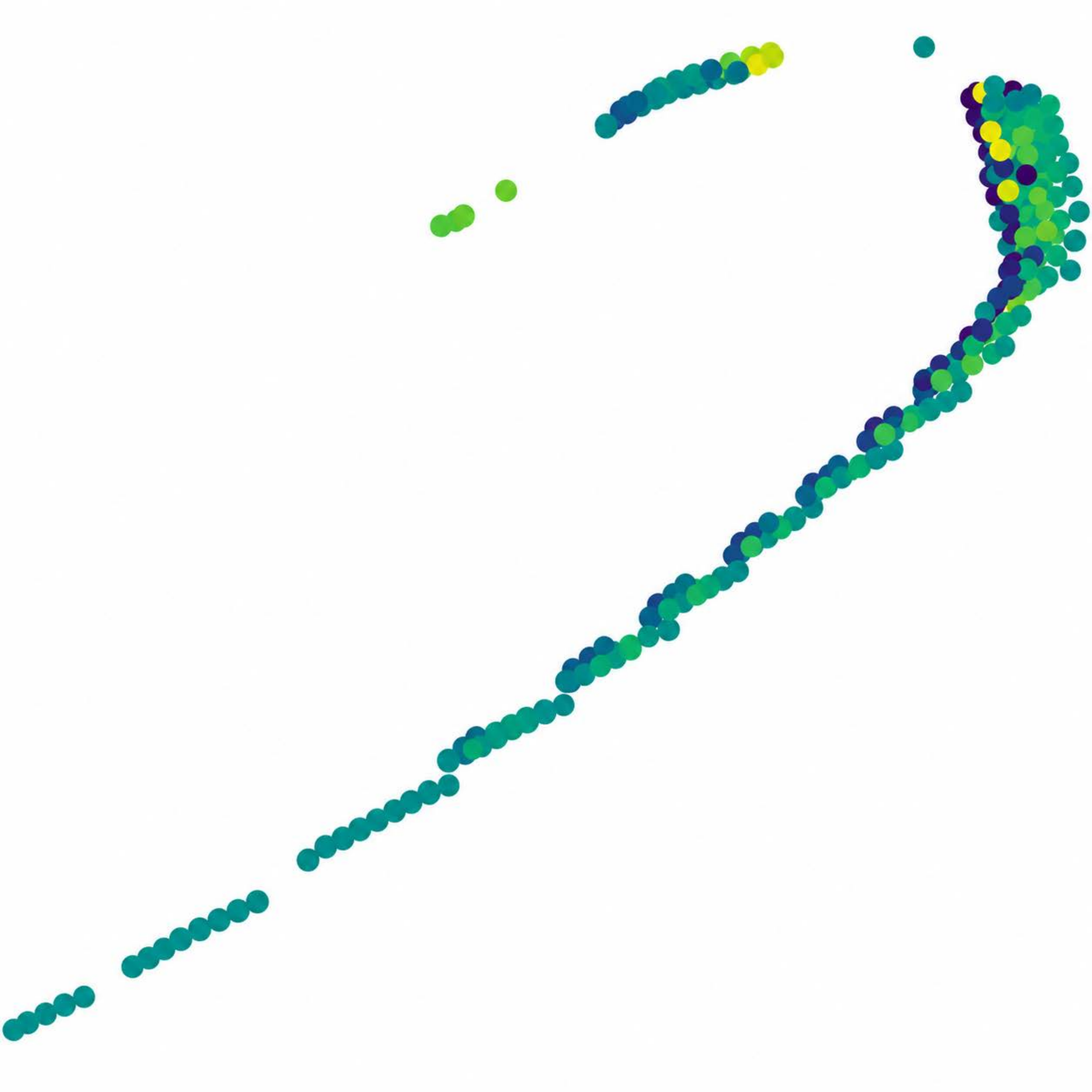}
		\caption{AE}
		\label{fig:scenario1_ae}
	\end{subfigure}
	\caption{Visualization of physical user locations and the corresponding  two-dimensional channel charts generated by different methods for Scenario~1. Colors indicate the physical locations of UEs.}
	\label{fig:scenario1_charts}
	\vspace{-5pt}
\end{figure*}

 Tables~\ref{tab:res1} and \ref{tab:res2} summarize the results for the two scenarios considered. Fig.~\ref{fig:scenario1_charts} further visualizes the physical UE locations and the corresponding two-dimensional charts produced by PACC and representative baselines in Scenario 1.

\begin{table}[htbp]
	\caption{Performance Comparison in Scenario 1}
	\label{tab:results1}
	\centering
	\renewcommand{\arraystretch}{1.0} % Compact row spacing.
	\begin{tabular}{@{}rccccc@{}} % @{} 用于去除表格左右两侧的多余空白
		\toprule
		\emph{Method} & TW$ \uparrow$ & CT$\uparrow$ & KS$\downarrow$ & RD$\downarrow$ & out-of-sample \\
		\midrule
		PCA & 0.8080 & 0.8789 & 0.7722 & 0.9526 & \Checkmark \\
		MDS & 0.7471 & 0.7484 & 0.5980 & 0.9877 & \XSolid \\
		UMAP & 0.9232 & 0.9247  & 0.4692 & 0.9600 & \XSolid \\
		t-SNE & 0.8561 & 0.9029 & 0.4353 & 0.9716 & \XSolid \\
		% Isomap & 0.8516 & 0.9175 & 0.4673 & 0.9501 & \Checkmark \\
		AE & 0.6906 & 0.8997 & 0.7485 & 0.9634 & \Checkmark \\
		GSN & 0.6849 & 0.8804 & 0.6111 & 0.9837 & \Checkmark \\
		% CS & 0.6216 & 0.7900 & 0.5622 & 0.9935 & \Checkmark \\
		ADP & 0.9164 & 0.9025 & 0.3806 & 0.9602 & \Checkmark \\
		% Siamese-ADP & 0.0 & 0.0 & 0.0 & 0.0 & easy \\
		\rowcolor{gray!20}
		PACC-$\mathcal{L}$ & \textbf{0.9597} & \textbf{0.9603} & \textbf{0.2922} & \textbf{0.8011} & \Checkmark \\
		PACC-$\mathcal{L}_{\text{tri}}$ & 0.8932 & 0.9229 & 0.4447 & 0.9452 & \Checkmark \\
		\bottomrule
	\end{tabular}
	\label{tab:res1}
\end{table}
\begin{table}[htbp]
	\caption{Performance Comparison in Scenario 2}
	\label{tab:results2}
	\centering
	\renewcommand{\arraystretch}{1.0} % Compact row spacing.
	\begin{tabular}{@{}rccccc@{}} % @{} 用于去除表格左右两侧的多余空白
		\toprule
		\emph{Method} & TW$ \uparrow$ & CT$\uparrow$ & KS$\downarrow$ & RD$\downarrow$ & out-of-sample \\
		\midrule
		PCA & 0.7746 & 0.8029 & 0.7955 & 0.9868 & \Checkmark \\
		MDS & 0.7419 & 0.7389 & 0.6886 & 0.9923 & \XSolid \\
		UMAP & 0.8991 & 0.9051  & 0.5104 & 0.9652 & \XSolid \\
		t-SNE & 0.8895 & 0.9065 & 0.4613 & 0.9613 & \XSolid \\
		% Isomap & 0.8516 & 0.9175 & 0.4673 & 0.9501 & \Checkmark \\
		AE & 0.7790 & 0.8622 & 0.6722 & 0.9720 & \Checkmark \\
		GSN & 0.7840 & 0.8467 & 0.6693 & 0.9832 & \Checkmark \\
		% CS & 0.6216 & 0.7900 & 0.5622 & 0.9935 & \Checkmark \\
		ADP & 0.8879 & 0.8617 & 0.5059 & 0.9721 & \Checkmark \\
		% Siamese-ADP & 0.0 & 0.0 & 0.0 & 0.0 & easy \\
		\rowcolor{gray!20}
		PACC-$\mathcal{L}$ & \textbf{0.9011} & \textbf{0.9176} & \textbf{0.4218} & \textbf{0.9089} & \Checkmark \\
		PACC-$\mathcal{L}_{\text{tri}}$ & 0.8839 & 0.8889 & 0.5039 & 0.9423 & \Checkmark \\
		\bottomrule
	\end{tabular}
	\label{tab:res2}
\end{table}

Specifically, in Scenario 1, PACC achieves the best performance across all four metrics, with TW, CT, KS, and RD values of 0.9597, 0.9603, 0.2922, and 0.8011, respectively. Compared with the strongest baseline, PACC improves TW and CT by 0.0365 and 0.0356, while reducing KS and RD by 0.0884 and 0.1515, respectively. These improvements indicate that PACC better preserves both neighborhood relationships and global channel geometry. This advantage is mainly attributed to the proposed propagation-aware constraint, where dominant-path ToA and AoA provide reliable geometric cues under LoS propagation.

Scenario 2 is more challenging because blockage and richer multipath propagation weaken the relationship between dominant-path parameters and UE locations. Nevertheless, PACC still achieves the best overall performance, with TW = 0.9011, CT = 0.9176, KS = 0.4218, and RD = 0.9089. Compared with the strongest baseline, it improves TW and CT by 0.0020 and 0.0111, while reducing KS and RD by 0.0395 and 0.0524, respectively. The consistent gains demonstrate the robustness of PACC under NLoS propagation. In this case, the ADM-based dissimilarity in (\ref{eq:nlosL}) effectively captures multipath similarities, providing meaningful geometric supervision when dominant-path information is unreliable.

We further evaluate the effect of the proposed physics-guided loss by comparing PACC with its triplet-only variant, PACC-$\mathcal{L}_{\mathrm{tri}}$. In Scenario 1, introducing the physics-guided constraint improves TW and CT by 0.0665 and 0.0374, while reducing KS and RD by 0.1525 and 0.1441, respectively. In Scenario 2, it improves TW and CT by 0.0172 and 0.0287, and reduces KS and RD by 0.0821 and 0.0334, respectively. These results confirm that triplet supervision can preserve neighborhood ordering but cannot fully determine the relative arrangement of neighboring samples. By explicitly constraining propagation-aware local geometry, the proposed loss substantially improves both local topology preservation and global geometric consistency. Moreover, unlike non-parametric approaches such as MDS and t-SNE, PACC learns an explicit forward mapping that can directly embed previously unseen CSI samples without reconstructing the entire channel chart.

\section{Conclusion}
We have proposed PACC, a propagation-aware channel charting framework that incorporates physical constraints into deep metric learning without requiring precise user locations or timestamp-based pseudo-labels. By extracting physically meaningful CSI features, including angle-delay information and channel gain, PACC constructs propagation-aware dissimilarities tailored to both LoS and NLoS conditions. Combined with the triplet-based neighborhood supervision, the proposed loss preserves both local neighborhood relationships and geometric consistency. Simulation results in representative LoS- and NLoS-dominant scenarios show that PACC consistently outperforms both conventional dimensionality-reduction baselines and existing learning-based channel charting approaches in terms of neighborhood preservation and global geometric fidelity. These results demonstrate the effectiveness of incorporating propagation knowledge into unsupervised CSI representation learning and highlight the potential of PACC for applications such as beam management and radio environment sensing.

\bibliographystyle{IEEEtran}
\begingroup
\scriptsize
\bibliography{ref}
\endgroup

\end{document}